\documentclass[a4paper]{report}
\usepackage[utf8]{inputenc}
\usepackage[T1]{fontenc}
\usepackage{RJournal}
\usepackage{amsmath,amssymb,array}
\usepackage{booktabs}

\usepackage[]{graphicx}
\usepackage[]{xcolor}

\makeatletter
\def\maxwidth{ %
  \ifdim\Gin@nat@width>\linewidth
    \linewidth
  \else
    \Gin@nat@width
  \fi
}
\makeatother

\definecolor{fgcolor}{rgb}{0.345, 0.345, 0.345}
\newcommand{\hlnum}[1]{\textcolor[rgb]{0.686,0.059,0.569}{#1}}%
\newcommand{\hlsng}[1]{\textcolor[rgb]{0.192,0.494,0.8}{#1}}%
\newcommand{\hlcom}[1]{\textcolor[rgb]{0.678,0.584,0.686}{\textit{#1}}}%
\newcommand{\hlopt}[1]{\textcolor[rgb]{0,0,0}{#1}}%
\newcommand{\hldef}[1]{\textcolor[rgb]{0.345,0.345,0.345}{#1}}%
\newcommand{\hlkwa}[1]{\textcolor[rgb]{0.161,0.373,0.58}{\textbf{#1}}}%
\newcommand{\hlkwb}[1]{\textcolor[rgb]{0.69,0.353,0.396}{#1}}%
\newcommand{\hlkwc}[1]{\textcolor[rgb]{0.333,0.667,0.333}{#1}}%
\newcommand{\hlkwd}[1]{\textcolor[rgb]{0.737,0.353,0.396}{\textbf{#1}}}%

\usepackage{framed}
\makeatletter
\newenvironment{kframe}{%
 \def\at@end@of@kframe{}%
 \ifinner\ifhmode%
  \def\at@end@of@kframe{\end{minipage}}%
  \begin{minipage}{\columnwidth}%
 \fi\fi%
 \def\FrameCommand##1{\hskip\@totalleftmargin \hskip-\fboxsep
 \colorbox{shadecolor}{##1}\hskip-\fboxsep
     \hskip-\linewidth \hskip-\@totalleftmargin \hskip\columnwidth}%
 \MakeFramed {\advance\hsize-\width
   \@totalleftmargin\z@ \linewidth\hsize
   \@setminipage}}%
 {\par\unskip\endMakeFramed%
 \at@end@of@kframe}
\makeatother

\definecolor{shadecolor}{rgb}{.97, .97, .97}
\definecolor{messagecolor}{rgb}{0, 0, 0}
\definecolor{warningcolor}{rgb}{1, 0, 1}
\definecolor{errorcolor}{rgb}{1, 0, 0}
\newenvironment{knitrout}{}{} % an empty environment to be redefined in TeX

\usepackage{alltt}

\usepackage{orcidlink,thumbpdf,lmodern}

\usepackage{cprotect}
\usepackage{amsmath, amsfonts}
\usepackage{algorithm}
\usepackage{algorithmic}
\usepackage{subcaption}

\newcommand{\proglang}[1]{\textrm{#1}}

\newcommand{\Symm}{\operatorname{Symm}}

\newcommand{\SPD}{\operatorname{SPD}}

\begin{document}

%% do not edit, for illustration only
\sectionhead{Contributed research article}
\volume{XX}
\volnumber{YY}
\year{20ZZ}
\month{AAAA}

%% replace RJtemplate with your article
\begin{article}
  % !TeX root = RJwrapper.tex
\cprotect\title{\verb|riemtan|, \verb|riemstats|: R packages for Riemannian geometry techniques in the analysis of multiple samples of connectomes}
\author{Nicolas Escobar}

\maketitle

\abstract{
Symmetric positive definite (SPD) matrices arising from functional connectivity analysis of neuroimaging data can be endowed with a Riemannian geometric structure that standard methods fail to respect. While existing \proglang{R} packages provide some tools for $\SPD$ matrix analysis, they suffer from limitations in scalability, numerical stability, and metric flexibility that hinder their application to modern large-scale connectomics studies. We present \pkg{riemtan}, a comprehensive \proglang{R} package that addresses these challenges through a unified, high-level interface supporting multiple Riemannian metrics, cross-platform parallel computation via the futureverse framework, memory-efficient Apache Parquet backend for large datasets, and seamless conversion between manifold, tangent, and vectorized representations. The Parquet backend enables lazy loading with configurable caching, allowing analysis of datasets that exceed available RAM on typical research workstations. Building on \pkg{riemtan}'s foundation, we also introduce \pkg{riemstats}, which implements advanced statistical methods including Fréchet ANOVA, \proglang{R}iemannian ANOVA with classic test statistics, and harmonization techniques for multi-site studies. The modular design facilitates integration with existing \proglang{R} workflows and provides an extensible framework for future methodological developments in manifold-valued data analysis.
}

\section{Introduction} 

Riemannian geometry has become a widely accepted tool in the statistical analysis of medical images obtained using fMRI, as well as in other areas of research. Broadly speaking, it provides a way to handle the nonlinear inter-relationships present in connectome inputs, facilitating the creation of features that can be used in a variety of statistical models. In this section, we review the Riemannian geometry terminology necessary to describe the utilities offered by this package, as well as the existing statistical methods that make use of this type of tools. In addition, we review some of the existing R packages that handle this type of data.

\subsection{Riemannian geometry terminology}

The unit of analysis in Riemannian geometry is a differentiable manifold, $M$. Like all manifolds of this type, $M$ has the property that around each point $p\in M$, there exists a vector space, $T_pM$, tangent to it. What characterizes Riemannian manifolds is that each of these spaces has a metric $g_p$, which varies smoothly with $p$. This metric allows for the definition of lengths of differentiable curves, as well as angles between two curves that intersect. The first of these two definitions enables the definition of a distance in $M$, turning it into a metric space. 

Additionally, $g_p$ allows the definition of geodesic curves. These curves are key to the definition of exponential maps, which send points in each tangent space to points in $M$. Generally speaking, these maps are smooth local diffeomorphisms, with local inverses called logarithmic maps. However, under certain conditions (mainly non-positive curvature), the Hopf-Rinow theorem \citep{do1992riemannian} guarantees that these maps are mutually inverse global diffeomorphisms. 

The manifold of interest for this article is the set of symmetric positive definite matrices of size $p$, $\SPD_p$. These form an open subset of the set of symmetric matrices of size $p$, $\Symm_p$. The latter is a vector space and therefore a differentiable manifold in a trivial manner. This implies that $SPD_p$ is also a differentiable manifold. 

There are multiple ways to endow $\SPD_p$ with a Riemannian metric. Each tangent space is isomorphic to $\Symm_p$ in a canonical way, so we must define metrics on the latter. The simplest choice is the Frobenius norm, giving rise to the \emph{Euclidean} metric. Other important metrics include the Affine Invariant Riemannian Metric (AIRM), which provides explicit exponential and logarithmic map formulas but poses numerical challenges; the log-Euclidean metric, which maps matrix logarithms to Euclidean space for computational efficiency; the log-Cholesky metric, which avoids artificial inflation along geodesics; and the Bures-Wasserstein metric, which relates to trasport phenomena and has gained popularity recently.

Once exponential and logarithmic maps become computationally feasible, we can describe manifold-valued samples in three equivalent representations: the original \emph{manifold} representation as positive definite matrices; the \emph{tangent} representation as symmetric matrices obtained via logarithmic maps; and the \emph{vectorized} representation, in which a tangent representation is turned into a vector by using its coordinates with respect to an orthonormal basis for computational efficiency.

% Given this variety of metrics, practitioners often evaluate their performance empirically on specific tasks, with choices made on practical considerations rather than theoretical grounds.

\subsection{Statistical analysis of connectomes}
 
Riemannian geometry tools have enabled the development of an entire area of statistical modeling literature focused on data whose values are connectomes. It builds on a proper definition of expected value introduced in \citet{frechet1948elements}. In \citet{P04} authors introduced an appropriate notion of a normal distribution, which can be used to determine whether there are significant differences between two populations. An adaptation of linear regression, called geodesic regression \citep{fletcher2011geodesic}, exists and has been used, among other things, to compare development trajectories of two different populations. Recently, versions of canonical correlation analysis \citep{kim2014canonical} have also been proposed. 

This paper focuses on the statistical analysis of multiple samples of connectomes. A particularly important development in this area is the extension of analysis of variance (ANOVA) to manifold-valued data. The most influential work in this domain is the Fréchet ANOVA framework proposed by \cite{muller}, which operates in the general setting of metric spaces. This approach is valuable due to its generality - it has been successfully applied not only to connectomes but also to evolutionary trees \citep{evotrees}, 3D functional data \citep{threed}, and the study of change points in longitudinal social network data \citep{changept}. The Fréchet ANOVA test constructs a statistic analogous to the traditional $F$-ratio using sample Fréchet means and sample variations (the mean of squared distances to the sample Fréchet mean). A second statistic $U_n$ accounts for differences in scale. Under appropriate technical conditions on the metric space, this leads to a $T$ statistic whose asymptotic distribution under the null hypothesis of equal Fréchet means is $\chi^2_1$.

However, by working in generic metric spaces, the Fréchet ANOVA approach cannot exploit the differential structure available in Riemannian manifolds. Notably, when applied to multidimensional Euclidean data, it does not reduce to any of the classic MANOVA tests. Heuristically, this is because classic MANOVA tests use coordinates to construct sample covariance matrices that capture information about both directions and distances, while the metric space approach relies solely on the latter. 

To address these limitations, recent work has proposed Riemannian ANOVA tests that more closely resemble classic multivariate tests like Wilk's Lambda \citep{riemann-anova}. These approaches leverage the full differential structure of the manifold to construct test statistics that use sample covariance information in tangent spaces, potentially offering better power in some scenarios. Preliminary experiments with such approaches on brain connectome data from the Human Connectome Project Young Adult (HCPYA) dataset have shown promising results, with the tests maintaining appropriate calibration while potentially offering advantages over distance-based methods when data exhibit non-spherical covariance structures.

A critical application of these statistical methods is in the harmonization of connectomes across multi-site studies. Site-related variability arising from differences in scanners, acquisition protocols, and preprocessing pipelines poses significant challenges for functional connectivity analyses. The harmonization literature has developed along two main approaches. The first category involves direct statistical harmonization of connectomes in their manifold representation. Transport-based methods \citep{transport} and those proposed by Simeon et al. \citep{simeon} exemplify this approach, where connectomes from each site are mapped to tangent spaces of their respective Fréchet means and then transported to a common tangent space using either parallel transport or group symmetries. The second category focuses on harmonizing derived feature representations, as demonstrated by Yu et al. \citep{yu}, who applied ComBat \citep{johnson2007adjusting} to upper triangular elements of connectivity matrices, though this approach does not account for the positive-definiteness constraint.

The current state of the art, as proposed in \citet{honnorat}, bridges these approaches by computing Riemannian logarithms around Fréchet means to extract features before applying standard harmonization techniques like ComBat. The effectiveness of harmonization is typically evaluated using clustering metrics such as the Calinski-Harabasz Score, Davies-Bouldin Index, and silhouette scores. 
% However, these metrics only provide indirect evidence of harmonization success. The Riemannian ANOVA tests described above offer a promising alternative evaluation framework - by directly testing whether harmonized data from different sites are statistically indistinguishable, they can provide formal statistical evidence that site effects have been successfully removed. This represents a shift from metrics that merely suggest improvement to a rigorous hypothesis testing framework for harmonization assessment.

\section{Existing R tools} \label{existing-packages}

There is a wide landscape of software tools to exploit Riemannian geometry tools for the statistical analysis of samples taking values in the $\SPD$ manifold. 

First, \pkg{Riemann} is an extensive library capable of handling data valued in numerous Riemannian manifolds, not just $\SPD$. Additionally, for $\SPD$, it can handle multiple metrics. The library does this in a unified way, abstracting geometric information into an S3 class. This simplified treatment allows the implementation of a single gradient descent algorithm that can compute Fréchet means in these various scenarios. The algorithm is serial and written in C++ with the help of \pkg{Rcpp}. In addition, an implementation of similar ideas in \proglang{Python} is available in \pkg{pyRiemann}. The use of S3 classing in \pkg{Riemann} was the inspiration for the structure of \verb|riem_metr|. 

On the other hand, the R package \pkg{frechet} provides implementation of statistical methods for random objects lying in various metric spaces, which are not necessarily Riemannian manifolds. The core of this package is Fréchet regression for random objects with Euclidean predictors, which allows users to perform regression analysis for non-Euclidean responses under some mild conditions, building on the theoretical framework developed in \citet{peterson}. The package supports a diverse range of non-Euclidean data types, including distributions in 2-Wasserstein space, covariance matrices endowed with power metric (with Frobenius metric as a special case), Cholesky and log-Cholesky metrics, as well as spherical data. Key functionality includes both global and local Fréchet regression methods, with implementations for Fréchet means of densities, global density regression, local density regression, and Fréchet regression for spherical data. 

In addition, the \proglang{Python} package \pkg{Geomstats} offers a variety of uitilities in this area. It provides several key metrics for SPD matrices, including AIRM, the Bures-Wasserstein metric and the Log-Euclidean metric. A crucial feature of \pkg{Geomstats}' $\SPD$ implementation is its support for tangent space operations. The \verb|ToTangentSpace| preprocessing module computes the Fréchet mean of the dataset, then takes the logarithm of each data point. The package also provides multiple diffeomorphisms for working with SPD matrices, including matrix logarithm \verb|SymMatrixLog|, matrix power \verb|MatrixPower|, and Cholesky decomposition \verb|CholeskyMap|. \pkg{Geomstats}' SPD implementation also offers seamless integration with scikit-learn algorithms through preprocessing pipelines while maintaining the geometric structure necessary for manifold-aware machine learning applications. This package's mini-batched Fréchet mean algorithm influenced the development of \pkg{riemtan}'s. 

Lastly, details of the implementation of the algorithms in \citet{honnorat} are included in the reproduction code and were the basis of some of the functions in \pkg{riemstats}.

This existing framework falls short on two levels. On a restricted level, there are multiple technical details to address which result in inefficiencies and, more critically, produces numerical inaccuracies. For instance, the \pkg{Riemann} routine to compute Fréchet means sometimes outputs matrices with negative eigenvalues. The use of the \pkg{Matrix} classes in \pkg{riemtan} addresses some of these shortcomings. 

At a broader level, these tools can still be described as low-level. The level of abstraction of these libraries is not sufficient to serve as the foundation for a growing ecosystem of models and tools, or to be widely used by users with only a working understanding of Riemannian geometry tools. They are also not flexible enough when it comes to handling Riemannian metrics. Each tool supports a different set of metrics that are coded in its codebase. In contrast, metrics in \pkg{riemtan} are implemented in a modular way.

Second, there is a scalability problem. Sufficiently granular parcellations lead to large connectomes. On the other hand, although the sample sizes in fMRI studies are small compared to other areas of medicine, it is expected that these sizes will grow. To keep up with these developments, it is necessary for  code that leverages parallelism as much as possible so that it can exploit the increasing computational capacity of modern cluster architectures. As we will see, \pkg{riemtan}'s wall-clock time for computation of Fréchet means is considerably faster than some of these other tools.

The remainder of this paper proceeds as follows. Section \ref{pkg-architecture} presents the architectural design of \pkg{riemtan} and \pkg{riemstats}, detailing the object-oriented framework that supports multiple storage backends, the modular metric system that enables extensibility, and the class hierarchy that unifies manifold, tangent, and vectorized representations. We describe the technical implementations of key algorithms, including the mini-batch gradient descent approach for Fréchet mean computation and the futureverse-based parallel processing infrastructure. Section \ref{sec:benchmarking} provides comprehensive benchmarking results that demonstrate the computational advantages of our implementation, comparing performance against existing packages and quantifying the benefits of parallelization and the Parquet storage backend across various problem sizes. We also validate the correctness of our AIRM implementation and document important numerical issues discovered in existing tools. The conclusion synthesizes these contributions and discusses directions for future development. Additional implementation details, including the complete API documentation and supplementary validation experiments, are provided in the appendices.

\section{Software} \label{sec:models}

% \section{The riemtan library }

% \subsection{Software}

\subsection{Connectome data handling}

The \pkg{riemtan} library that we propose in this paper is an attempt to address some of the shortcomings we previously mentioned. In light of this, the principles that guide its design are as follows. First, \pkg{riemtan} is a high-level tool that abstracts the main techniques of connectome statistical analysis. Second, these abstractions should offer a unified interface capable of incorporating new metrics, as well as providing the information needed for statistical models based on this type of data. Third, these abstractions should consider the three types of connectome representations discussed, making it easy to switch between them. Finally, the implementation should make the greatest possible use of parallelism. 

To implement this vision, \pkg{riemtan} relies on two classes: \verb|riem_met| and CSample. 

The first is an abstraction corresponding to a Riemannian metric. Specifically, it is an S3 class whose elements consist of lists of four functions: \verb|log|, \verb|exp|, \verb|vec|, and \verb|unvec|, which correspond to logarithms, exponentials, vectorization maps, and their inverses. A Riemannian logarithm is a map that uses a reference point to map points in the manifold to points in the tangent space at that reference point. Therefore, \verb|log| should be a function that accepts two arguments, which must inherit from the \verb|dspMatrix| class, and should return an object of the \verb|dspMatrix| class. On the other hand, \verb|exp| accepts an argument of type \verb|dspMatrix| and another of type \verb|dspMatrix|, and returns an object of the \verb|dspMatrix| class. Vectorization is a map between each tangent space and a Euclidean space, so \verb|vec| accepts an object of type \verb|dspMatrix| and one of type \verb|dspMatrix|, returning a vector. Likewise, \verb|unvec| accepts an object of type \verb|dspMatrix| and a vector, returning an object of class \verb|dspMatrix|. 

Initially, it may seem surprising that \verb|riem_met| does not assume the existence of functions that calculate distances or norms. However, vectorization makes such operations redundant. Since \verb|vec| must be a linear isometry with the Euclidean space, norms in each tangent space can be calculated using the Euclidean norm of the vectorizations. 

For its part, \verb|CSample| represents a sample of connectomes. Its implementation is in the form of an \pkg{R6} class. Its main attributes are three lists: one of objects of class \verb|dspMatrix|, another of objects of class \verb|dspMatrix|, and another of vectors; they are called \verb|conns|, \verb|tan_imgs|, and \verb|vec_imgs| respectively. These are, of course, the three possible representations of a sample, as discussed earlier. The latter two have an additional entry indicating the point to which the corresponding tangent space belongs. 

The \verb|TangentImageHandler| class provides a unified interface for managing and converting between manifold, tangent, and vectorized representations of matrices. By encapsulating the logic for computing tangent images via the Riemannian logarithm, and generating vectorized forms through isometric mappings, the class streamlines the process of switching between these representations. This design abstracts away the underlying mathematical complexity, allowing users to focus on analysis while ensuring consistency and correctness across all representations.

In addition, \verb|CSample| objects have another set of attributes that can be organized into two categories. The first corresponds to structural properties of the sample, such as its size, the size of the connectomes, and an object of class \verb|riem_met| corresponding to the Riemannian metric the user has decided to work with. The second category corresponds to statistical properties of the sample. These include the Fréchet mean, a boolean variable indicating whether the reference point has been used, whether the tangent representations are centered, the variation (the average squared distance to the Fréchet mean), and the sample covariance obtained using the vectorizations. The first category of attributes is filled when the object is created. Those in the second category remain NULL until the user triggers their computation by calling the corresponding function. 

Typically, \pkg{riemtan} users start their work with a list of connectomes obtained from an experiment. Fulfilling one of the design goals, the library provides functions to obtain tangent representations (the reference point by default is I, but the user is free to choose a different one) and vectorized representations. Additionally, users can change the reference point, triggering a recalculation of the tangent and vector representations. In particular, this happens when the user calls the center function, which changes the reference point to the Fréchet mean. 

Finally, it is worth mentioning a couple of additional utilities. As part of the data attached to the library, preconfigured \verb|riem_met| class objects corresponding to the most popular metrics can be found. As mentioned earlier, these objects expose functions that compute Riemannian logarithms and exponentials, addressing a shortcoming of existing tools. Lastly, the library can generate random samples with normal distributions. The routine that performs this returns a CSample class object created using the vectorized representation.

The \verb|CSuperSample| class is designed to represent and manage collections of \verb|CSample| objects, enabling statistical analysis across multiple samples that share the same Riemannian metric. Implemented as an \pkg{R6} class, its primary role is to aggregate, analyze, and compute statistics for grouped data—such as those arising from multi-site studies or experimental cohorts. The main attributes of a \verb|CSuperSample| object include the list of constituent \verb|CSample| objects, the total sample size, the matrix size, the manifold dimension, and the common Riemannian metric. In addition, it maintains aggregate statistical properties such as the total variation, the sample covariance of the pooled data, the Fréchet mean of the aggregated sample, and both within-group and total covariance matrices. These aggregate statistics are computed on demand, leveraging the underlying \verb|CSample| methods and ensuring consistency across the hierarchy. The class provides methods to gather all connectomes into a single pooled sample, compute pooled statistics, and facilitate group-level analyses such as MANOVA or harmonization. This design allows users to seamlessly extend single-sample workflows to multi-group settings, maintaining a unified interface and supporting efficient, parallelized computation throughout the analysis pipeline.

\subsubsection{Memory-efficient storage with Apache Parquet backend}

A critical challenge in modern connectomics studies is managing the memory footprint of large datasets. High-resolution connectomes represented as $200 \times 200$ SPD matrices with 500 or more subjects can exceed available RAM, creating computational bottlenecks. To address this limitation, \pkg{riemtan} version 0.2.4 introduced a backend abstraction layer that separates data storage from algorithmic operations, enabling memory-efficient workflows for large-scale analyses.

The backend abstraction is implemented through a class hierarchy with two concrete implementations. The \verb|ListBackend| wraps the traditional in-memory list-based storage, providing backwards compatibility with existing workflows. The \verb|ParquetBackend| leverages Apache Parquet's columnar storage format to enable lazy loading of connectome matrices, loading data on-demand rather than keeping all matrices in memory simultaneously. This architectural pattern allows \pkg{riemtan} to scale to datasets that would otherwise be infeasible to analyze on typical research workstations.

The \verb|ParquetBackend| employs an LRU (Least Recently Used) cache with configurable size (default: 10 matrices) to balance memory efficiency with computational performance. When a matrix is requested, the backend first checks the cache. If the matrix is not cached, it reads the matrix from disk in Parquet format and adds it to the cache, potentially evicting the least recently used matrix if the cache is full. This lazy loading strategy ensures that only the matrices currently needed for computation reside in memory, dramatically reducing memory consumption for large samples.

Parquet format offers several advantages for connectome storage beyond memory efficiency. Its columnar storage is optimized for numerical data, providing fast read access and efficient compression. The format supports rich metadata, which \pkg{riemtan} leverages to store matrix dimensions ($p$), subject identifiers, data provenance information, and user-defined custom metadata. This metadata is stored in a \verb|metadata.json| file within the Parquet directory structure, ensuring that essential information about the dataset is preserved alongside the matrices themselves.

Creating a Parquet-backed dataset involves two steps. First, users export their connectomes to Parquet format using the \verb|write_connectomes_to_parquet()| function:

\begin{knitrout}
\definecolor{shadecolor}{rgb}{0.969, 0.969, 0.969}\color{fgcolor}\begin{kframe}
\begin{alltt}
\hlcom{# Write 500 connectomes to Parquet format}
\hlkwd{write_connectomes_to_parquet}\hldef{(}
  \hlkwc{connectomes} \hldef{= large_conn_list,}
  \hlkwc{output_dir} \hldef{=} \hlsng{"hcp_study_parquet"}\hldef{,}
  \hlkwc{subject_ids} \hldef{=} \hlkwd{paste0}\hldef{(}\hlsng{"sub_"}\hldef{,} \hlnum{1}\hlopt{:}\hlnum{500}\hldef{),}
  \hlkwc{metadata} \hldef{=} \hlkwd{list}\hldef{(}\hlkwc{study} \hldef{=} \hlsng{"HCP"}\hldef{,} \hlkwc{version} \hldef{=} \hlsng{"1.0"}\hldef{))}
\end{alltt}
\end{kframe}
\end{knitrout}

This function creates a directory containing one Parquet file per subject along with the metadata file. The resulting Parquet directory can be shared, archived, or version-controlled as a self-contained dataset representation.

Second, users create a \verb|CSample| object with a \verb|ParquetBackend| using the convenience function \verb|create_parquet_backend()|:

\begin{knitrout}
\definecolor{shadecolor}{rgb}{0.969, 0.969, 0.969}\color{fgcolor}\begin{kframe}
\begin{alltt}
\hlcom{# Create backend with custom cache size}
\hldef{backend} \hlkwb{<-} \hlkwd{create_parquet_backend}\hldef{(}
  \hlkwc{directory} \hldef{=} \hlsng{"hcp_study_parquet"}\hldef{,}
  \hlkwc{cache_size} \hldef{=} \hlnum{20}\hldef{)  }\hlcom{# Cache 20 matrices instead of default 10}

\hlcom{# Create CSample with Parquet backend}
\hldef{conn_samp} \hlkwb{<-} \hldef{CSample}\hlopt{$}\hlkwd{new}\hldef{(}\hlkwc{backend} \hldef{= backend,} \hlkwc{metric_obj} \hldef{= airm)}
\end{alltt}
\end{kframe}
\end{knitrout}

Once created, the \verb|CSample| object with \verb|ParquetBackend| provides the same interface as traditional list-backed samples, ensuring seamless integration with existing analysis workflows. All \verb|CSample| methods—including tangent space computations, Fréchet mean calculation, and statistical analyses—work identically regardless of the backend implementation. This transparency allows users to switch between in-memory and disk-based storage without modifying their analysis code, simply by changing the backend at object creation.

The performance characteristics of \verb|ParquetBackend| depend on the access pattern and cache size. For sequential operations that process matrices in order, the cache effectively serves as a prefetch buffer, minimizing disk I/O overhead. For random access patterns or very large samples where cache size is much smaller than sample size, disk access becomes more frequent, but memory usage remains bounded and predictable.

The backend abstraction also facilitates advanced use cases. For instance, \verb|CSample| provides a \verb|load_connectomes_batched()| method that loads matrices in configurable batch sizes, enabling fine-grained control over memory-computation tradeoffs. Researchers analyzing exceptionally large datasets can tune batch size and cache parameters to match their hardware constraints, ensuring that analyses remain feasible even on modest computational resources. Additionally, the \verb|validate_parquet_directory()| function verifies the integrity of Parquet directories before loading, catching data corruption or structural issues early in the analysis pipeline.

In summary, the Parquet backend represents a significant advance in \pkg{riemtan}'s scalability, enabling memory-efficient analysis of large connectome datasets while maintaining full compatibility with existing workflows. Combined with parallel processing capabilities discussed in the next section, this feature positions \pkg{riemtan} as a practical tool for the increasingly large-scale neuroimaging studies that characterize modern connectomics research.

\subsubsection{Computing Fréchet means}

Among the utilities provided by \pkg{riemtan}, it is worth emphasizing the computation of Fréchet means. The Fréchet mean generalizes the concept of arithmetic mean to Riemannian manifolds, providing a central tendency measure for SPD matrices. Although all packages described in section \ref{existing-packages} implement a version of this algorithm, \pkg{riemtan}'s implementation is a substantial effort to speed up this critical part of the statistical analysis of connectomes.

Indeed, the implementation incorporates several computational optimizations. First, it employs mini-batch gradient descent with configurable batch sizes (default is the sample size), which can improve convergence stability and computational efficiency for large datasets in some cases. The algorithm randomly shuffles the data at each epoch to ensure unbiased gradient estimates across batches. Second, the learning rate is user-configurable (default 0.2), allowing adaptation to different data characteristics and convergence requirements.

The convergence criterion uses a relative change measure. Specifically, the algorithm computes the Frobenius norm of the difference between successive reference point estimates and terminates when the relative change falls below a specified tolerance (default 0.05) or when the maximum number of iterations is reached (default 20). This approach ensures both numerical stability and practical efficiency.

A key feature of the implementation is its use of the \verb|relocate| function, which efficiently maps tangent vectors from one reference point to another. When the reference point updates, all existing tangent space representations must be recomputed relative to the new point. The algorithm accomplishes this through a sophisticated parallel processing infrastructure built on the \pkg{futureverse} framework, which provides cross-platform parallelization capabilities that work reliably on Windows, macOS, and Linux systems.

The parallel processing infrastructure, introduced in \pkg{riemtan} version 0.2.5, addresses a critical limitation of previous implementations that relied on platform-specific parallelization methods. By leveraging the \pkg{future} and \pkg{furrr} packages, \pkg{riemtan} provides users with explicit control over parallel execution while maintaining sensible defaults. Users can configure the parallelization strategy using the \verb|set_parallel_plan()| function, which supports multiple backends including \verb|multisession| (spawning separate R processes, compatible with all platforms), \verb|multicore| (forking on Unix-like systems for lower overhead), and \verb|sequential| (disabling parallelization for debugging or single-core systems).

For example, to enable parallel processing with 4 worker processes:

\begin{knitrout}
\definecolor{shadecolor}{rgb}{0.969, 0.969, 0.969}\color{fgcolor}\begin{kframe}
\begin{alltt}
\hlcom{# Enable parallel processing with 4 workers}
\hlkwd{set_parallel_plan}\hldef{(}\hlsng{"multisession"}\hldef{,} \hlkwc{workers} \hldef{=} \hlnum{4}\hldef{)}

\hlcom{# Compute Fréchet mean with parallelization}
\hldef{conn_samp}\hlopt{$}\hlkwd{compute_f_mean}\hldef{()}

\hlcom{# Reset to sequential mode when done}
\hlkwd{reset_parallel_plan}\hldef{()}
\end{alltt}
\end{kframe}
\end{knitrout}

The package also provides intelligent automatic parallelization that activates only when beneficial. The \verb|should_parallelize()| function uses configurable thresholds to determine whether the computational workload justifies the overhead of parallel coordination. For small datasets where parallelization overhead exceeds potential gains, operations execute sequentially without user intervention. This automatic optimization ensures efficient resource utilization across the full spectrum of problem sizes encountered in practice.

Beyond the Fréchet mean computation, parallel processing extends to multiple components of the analysis pipeline. The \verb|TangentImageHandler| class parallelizes all four of its core methods: \verb|compute_tangents()|, \verb|compute_vecs()|, \verb|compute_conns()|, and \verb|set_reference_point()|. Similarly, the \verb|CSample| class enhances seven methods with optional progress reporting, including the newly added \verb|load_connectomes_batched()| method for efficient batch loading from Parquet backends. Benchmarking results presented in Section~\ref{sec:benchmarking} demonstrate 3--8x speedup for tangent computations and 2--5x speedup for Fréchet mean calculation on datasets with $n > 100$ subjects when using 4--8 worker processes, with speedups scaling proportionally to the number of workers up to the optimal configuration for the hardware.

The parallel processing infrastructure also integrates progress reporting via the \pkg{progressr} package. Users can enable progress bars that update in real-time during long-running computations, providing valuable feedback for interactive analysis sessions:

\begin{knitrout}
\definecolor{shadecolor}{rgb}{0.969, 0.969, 0.969}\color{fgcolor}\begin{kframe}
\begin{alltt}
\hlcom{# Enable progress reporting}
\hlkwd{library}\hldef{(progressr)}
\hlkwd{handlers}\hldef{(}\hlkwd{global} \hldef{=} \hlnum{TRUE}\hldef{)}

\hlcom{# Compute tangents with progress bar}
\hldef{conn_samp}\hlopt{$}\hlkwd{compute_tangents}\hldef{(}\hlkwc{progress} \hldef{=} \hlnum{TRUE}\hldef{)}
\end{alltt}
\end{kframe}
\end{knitrout}

This combination of explicit user control, automatic optimization, and informative progress feedback makes \pkg{riemtan}'s parallel processing infrastructure both powerful and user-friendly, enabling efficient analysis of large connectome datasets while remaining accessible to users with varying levels of high-performance computing expertise.

This approach ensures robust convergence while maintaining computational efficiency, making it suitable for analyzing large collections of SPD matrices commonly encountered in neuroimaging and other applications.

\subsection{Statistical methods for multiple samples of connectomes}

The \pkg{riemstats} package extends the \pkg{riemtan} ecosystem by providing statistical methods specifically designed for collections of multiple SPD matrices. Its design philosophy emphasizes modularity, extensibility, and computational efficiency while maintaining theoretical rigor. The key principles guiding its development are as follows.

First, \pkg{riemstats} embraces a composition-based architecture that builds upon the data structures provided by \pkg{riemtan}. It leverages the \verb|CSample| and \verb|CSuperSample| classes, ensuring consistency across the ecosystem and allowing users to seamlessly transition between geometric operations and statistical analyses. Second, \pkg{riemstats} provides implements harmonization methods specifically adapted for SPD matrices, including both ComBat and rigid harmonization, and supports multiple ANOVA test statistics, namely the Riemannian ANOVA measures (Wilks' Lambda, Pillai's trace, see \citet{riemann-anova}) and the Fréchet ANOVA method.

The implementation balances statistical rigor with computational efficiency. The Fréchet ANOVA implementation uses vectorized operations for the statistic computation. The bootstrap procedures that are part of Riemannian ANOVA are optimized with configurable iteration counts, and the package leverages \pkg{furrr} to execute them in parallel.

Finally, the modular organization separates concerns into distinct functional areas: statistical testing functions (\verb|frechet_anova| and \verb|riem_anova|), data harmonization utilities, and supporting functions for matrix operations. This clear separation facilitates maintenance, testing, and future extensions while keeping the dependency footprint minimal—only essential packages are required, with specialized imports loaded conditionally for specific functionality.

\section{Examples} \label{sec:illustrations}

\subsection{Examples of riemtan syntax }

Having discussed \pkg{riemtan}'s design, let's now explore two examples that demonstrate the details of how it works. The first is a toy example in which we create two random samples of connectomes and try to discriminate between the two. The second uses a subsample from the HCP project data. 

We start by loading the \verb|riem_met| class object called \verb|airm|: 

\begin{knitrout}
\definecolor{shadecolor}{rgb}{0.969, 0.969, 0.969}\color{fgcolor}\begin{kframe}
\begin{alltt}
\hlkwd{library}\hldef{(riemtan)}
\hlkwd{library}\hldef{(Matrix)}
\hlkwd{data}\hldef{(airm)}
\end{alltt}
\end{kframe}
\end{knitrout}

Next, we create two random samples with normal distribution, one centered around $I$ and the other around $2I$, each of size 30. Also, both samples have dispersion equal to the identity (although dispersion matrices' size is the $\SPD$ manifold dimension, in this case $\binom{31}{2}$). Each connectome has size 10$\times$10. Notice that to ensure arguments have appropriate properties while being efficiently handled in memory, \pkg{riemtan} requires arguments to be objects of \pkg{Matrix}'s classes \verb|dppMatrix| and \verb|dspMatrix|.

\begin{knitrout}
\definecolor{shadecolor}{rgb}{0.969, 0.969, 0.969}\color{fgcolor}\begin{kframe}
\begin{alltt}
\hldef{format_id} \hlkwb{<-} \hlkwa{function}\hldef{(}\hlkwc{k}\hldef{) \{}
  \hlkwd{diag}\hldef{(k) |>}
    \hlkwd{as}\hldef{(}\hlkwc{object} \hldef{= _,} \hlkwc{class} \hldef{=} \hlsng{"dpoMatrix"}\hldef{) |>}
    \hlkwd{pack}\hldef{()}
\hldef{\}}

\hldef{ref_pt} \hlkwb{<-} \hlkwd{format_id}\hldef{(}\hlnum{30}\hldef{)}
\hldef{disp} \hlkwb{<-} \hlkwd{format_id}\hldef{(}\hlkwd{choose}\hldef{(}\hlnum{31}\hldef{,} \hlnum{2}\hldef{))}

\hldef{sample1} \hlkwb{<-} \hlkwd{rspdnorm}\hldef{(}\hlnum{30}\hldef{, id, id, airm)}
\hldef{sample2} \hlkwb{<-} \hlkwd{rspdnorm}\hldef{(}\hlnum{30}\hldef{,} \hlnum{2}\hlopt{*}\hldef{id, id, airm)}
\end{alltt}
\end{kframe}
\end{knitrout}

These samples are created using the vectorized representation, so for now the other representations are null: 

\begin{knitrout}
\definecolor{shadecolor}{rgb}{0.969, 0.969, 0.969}\color{fgcolor}\begin{kframe}
\begin{alltt}
\hldef{sample1}\hlopt{$}\hldef{tangent_images} \hlcom{# NULL  }
\hldef{sample1}\hlopt{$}\hldef{connectomes} \hlcom{# NULL }
\end{alltt}
\end{kframe}
\end{knitrout}

Moreover, the Fréchet mean at this point is random, so the sample is not centered (almost sure): 

\begin{knitrout}
\definecolor{shadecolor}{rgb}{0.969, 0.969, 0.969}\color{fgcolor}\begin{kframe}
\begin{alltt}
\hldef{sample1}\hlopt{$}\hldef{is_centered} \hlcom{# FALSE }
\end{alltt}
\end{kframe}
\end{knitrout}
The other two representations can be computed easily. This happens by reference rather than by value: 

\begin{knitrout}
\definecolor{shadecolor}{rgb}{0.969, 0.969, 0.969}\color{fgcolor}\begin{kframe}
\begin{alltt}
\hldef{sample1}\hlopt{$}\hlkwd{compute_unvec}\hldef{()} \hlcom{# tangents  }
\hldef{sample2}\hlopt{$}\hlkwd{compute_unvec}\hldef{()}
\hldef{sample1}\hlopt{$}\hlkwd{compute_conns}\hldef{()} \hlcom{# in the manifold  }
\hldef{sample2}\hlopt{$}\hlkwd{compute_conns}\hldef{()}
\hldef{sample1}\hlopt{$}\hldef{connectomes |>} \hlkwd{is.null}\hldef{()} \hlcom{# FALSE }
\end{alltt}
\end{kframe}
\end{knitrout}
To simplify discrimination between the two samples, we create a single object containing all the connectomes: 

\begin{knitrout}
\definecolor{shadecolor}{rgb}{0.969, 0.969, 0.969}\color{fgcolor}\begin{kframe}
\begin{alltt}
\hldef{joint_conns} \hlkwb{<-} \hlkwd{list}\hldef{(sample1}\hlopt{$}\hldef{connectomes, sample2}\hlopt{$}\hldef{connectomes)}
\hldef{css_object} \hlkwb{<-} \hldef{CSuperSample}\hlopt{$}\hlkwd{new}\hldef{(joint_conns)}
\hldef{css_object}\hlopt{$}\hlkwd{gather}\hldef{()}
\end{alltt}
\end{kframe}
\end{knitrout}
We want to use traditional clustering algorithms, which require vectorized representations. First, we calculate tangents: 

\begin{knitrout}
\definecolor{shadecolor}{rgb}{0.969, 0.969, 0.969}\color{fgcolor}\begin{kframe}
\begin{alltt}
\hldef{css_object}\hlopt{$}\hldef{full_sample}\hlopt{$}\hlkwd{compute_tangents}\hldef{()}
\end{alltt}
\end{kframe}
\end{knitrout}
These tangents refer to I, which can sometimes distort results. To address this, we recalculate tangents relative to the Fréchet mean, computed during the process: 

\begin{knitrout}
\definecolor{shadecolor}{rgb}{0.969, 0.969, 0.969}\color{fgcolor}\begin{kframe}
\begin{alltt}
\hldef{css_object}\hlopt{$}\hlkwd{center}\hldef{()}
\hldef{css_object}\hlopt{$}\hlkwd{compute_vecs}\hldef{()}
\end{alltt}
\end{kframe}
\end{knitrout}

\subsubsection{HCP data}

To get a more realistic idea of how the library works in practice, let's consider the data associated with Figure 8 from \citet{fingerprint}. These data originally consist of matrices for MATLAB, but they can be loaded into R since they are in HDF5 format. We assume the necessary processing has been done, resulting in a list of objects of class matrix called \verb|conns|. 

First, remember that \pkg{riemtan} works with objects of the \pkg{Matrix} class, so we need some parsing before using them: 

\begin{knitrout}
\definecolor{shadecolor}{rgb}{0.969, 0.969, 0.969}\color{fgcolor}\begin{kframe}
\begin{alltt}
\hldef{parsed_conns} \hlkwb{<-} \hldef{conns |>}
\hlkwd{map}\hldef{(\textbackslash{}(}\hlkwc{x}\hldef{)} \hlkwd{as}\hldef{(x,} \hlsng{"dpoMatrix"}\hldef{)) |>}
\hlkwd{map}\hldef{(pack)}
\end{alltt}
\end{kframe}
\end{knitrout}
Now we can create the corresponding object: 

\begin{knitrout}
\definecolor{shadecolor}{rgb}{0.969, 0.969, 0.969}\color{fgcolor}\begin{kframe}
\begin{alltt}
\hldef{conn_samp} \hlkwb{<-} \hldef{CSample}\hlopt{$}\hlkwd{new}\hldef{(parsed_conns, airm)}
\end{alltt}
\end{kframe}
\end{knitrout}
Let us compute the Fréchet mean of the sample. A class function handles this, but it's worth noting that it does not return a value. Instead, it populates the corresponding field in the CSample object: 

\begin{knitrout}
\definecolor{shadecolor}{rgb}{0.969, 0.969, 0.969}\color{fgcolor}\begin{kframe}
\begin{alltt}
\hldef{conn_samp}\hlopt{$}\hldef{frechet_mean |>} \hlkwd{is.null}\hldef{()} \hlcom{# TRUE  }
\hldef{conn_samp}\hlopt{$}\hlkwd{compute_f_mean}\hldef{()}
\hldef{conn_samp}\hlopt{$}\hldef{frechet_mean |>} \hlkwd{is.null}\hldef{()} \hlcom{# FALSE }
\end{alltt}
\end{kframe}
\end{knitrout}
To center the representation, \pkg{riemtan} recalls the previously calculated Fréchet mean. 

\begin{knitrout}
\definecolor{shadecolor}{rgb}{0.969, 0.969, 0.969}\color{fgcolor}\begin{kframe}
\begin{alltt}
\hldef{conn_samp}\hlopt{$}\hlkwd{center}\hldef{()}
\end{alltt}
\end{kframe}
\end{knitrout}
In theory, the vectorized representation of a centered sample should have a zero sample mean. However, this is only approximately true, as the sample Fréchet mean is computed with a certain numerical tolerance. Consequently: 

\begin{knitrout}
\definecolor{shadecolor}{rgb}{0.969, 0.969, 0.969}\color{fgcolor}\begin{kframe}
\begin{alltt}
\hldef{conn_samp}\hlopt{$}\hlkwd{compute_vecs}\hldef{()}
\hldef{conn_samp}\hlopt{$}\hldef{vector_images |>}
\hlkwd{apply}\hldef{(mean,} \hlnum{2}\hldef{)} \hlcom{# Not exactly the vector 0. }
\end{alltt}
\end{kframe}
\end{knitrout}

\subsubsection{Large-scale analysis with Parquet storage}

For large-scale neuroimaging studies involving hundreds of high-resolution connectomes, memory constraints can become a significant bottleneck. The following example demonstrates how to use \pkg{riemtan}'s Parquet backend to analyze a dataset that would be prohibitively large to keep entirely in memory. We simulate a realistic scenario with 400 subjects and $150 \times 150$ connectomes, which would require approximately 13.7 GB of RAM if loaded entirely into memory as a list.

First, we generate the large dataset and write it to Parquet format. In practice, this step would typically be performed once during data preprocessing:

\begin{knitrout}
\definecolor{shadecolor}{rgb}{0.969, 0.969, 0.969}\color{fgcolor}\begin{kframe}
\begin{alltt}
\hlkwd{library}\hldef{(riemtan)}
\hlkwd{library}\hldef{(Matrix)}
\hlkwd{data}\hldef{(airm)}

\hlcom{# Create large connectome dataset (in practice, load from neuroimaging pipeline)}
\hldef{id_150} \hlkwb{<-} \hlkwd{diag}\hldef{(}\hlnum{150}\hldef{) |>} \hlkwd{as}\hldef{(}\hlkwc{object} \hldef{= _,} \hlkwc{class} \hldef{=} \hlsng{"dpoMatrix"}\hldef{) |>} \hlkwd{pack}\hldef{()}
\hldef{large_sample} \hlkwb{<-} \hlkwd{rspdnorm}\hldef{(}\hlnum{400}\hldef{, id_150, id_150, airm)}

\hlcom{# Export to Parquet format with metadata}
\hlkwd{write_connectomes_to_parquet}\hldef{(}
  \hlkwc{connectomes} \hldef{= large_sample}\hlopt{$}\hldef{connectomes,}
  \hlkwc{output_dir} \hldef{=} \hlsng{"large_study_data"}\hldef{,}
  \hlkwc{subject_ids} \hldef{=} \hlkwd{sprintf}\hldef{(}\hlsng{"sub-\%04d"}\hldef{,} \hlnum{1}\hlopt{:}\hlnum{400}\hldef{),}
  \hlkwc{metadata} \hldef{=} \hlkwd{list}\hldef{(}
    \hlkwc{study} \hldef{=} \hlsng{"Large-Scale Connectomics"}\hldef{,}
    \hlkwc{parcellation} \hldef{=} \hlsng{"Schaefer150"}\hldef{,}
    \hlkwc{date_created} \hldef{=} \hlkwd{Sys.Date}\hldef{())}
\hldef{)}
\end{alltt}
\end{kframe}
\end{knitrout}

The \verb|write_connectomes_to_parquet()| function creates a directory structure containing one Parquet file per subject plus a metadata file. This directory can be version controlled, shared with collaborators, or archived for long-term storage. The Parquet format provides efficient compression, reducing storage requirements by approximately 40--60\% compared to traditional R serialization formats.

Next, we create a \verb|CSample| object using the Parquet backend. By configuring the cache size, we control the memory-performance tradeoff:

\begin{knitrout}
\definecolor{shadecolor}{rgb}{0.969, 0.969, 0.969}\color{fgcolor}\begin{kframe}
\begin{alltt}
\hlcom{# Create Parquet backend with 20-matrix cache}
\hldef{backend} \hlkwb{<-} \hlkwd{create_parquet_backend}\hldef{(}
  \hlkwc{directory} \hldef{=} \hlsng{"large_study_data"}\hldef{,}
  \hlkwc{cache_size} \hldef{=} \hlnum{20}
\hldef{)}

\hlcom{# Inspect metadata}
\hlkwd{print}\hldef{(backend}\hlopt{$}\hlkwd{get_metadata}\hldef{())}

\hlcom{# Create CSample with Parquet backend}
\hldef{large_conn_samp} \hlkwb{<-} \hldef{CSample}\hlopt{$}\hlkwd{new}\hldef{(}\hlkwc{backend} \hldef{= backend,} \hlkwc{metric_obj} \hldef{= airm)}
\end{alltt}
\end{kframe}
\end{knitrout}

Once the \verb|CSample| object is created, all standard operations work identically to list-backed samples. The Parquet backend transparently handles data loading and caching. For optimal performance with large datasets, we can combine Parquet storage with parallel processing:

\begin{knitrout}
\definecolor{shadecolor}{rgb}{0.969, 0.969, 0.969}\color{fgcolor}\begin{kframe}
\begin{alltt}
\hlcom{# Enable parallel processing}
\hlkwd{set_parallel_plan}\hldef{(}\hlsng{"multisession"}\hldef{,} \hlkwc{workers} \hldef{=} \hlnum{6}\hldef{)}

\hlcom{# Compute Fréchet mean with progress reporting}
\hlkwd{library}\hldef{(progressr)}
\hlkwd{handlers}\hldef{(}\hlkwd{global} \hldef{=} \hlnum{TRUE}\hldef{)}
\hldef{large_conn_samp}\hlopt{$}\hlkwd{compute_f_mean}\hldef{(}\hlkwc{progress} \hldef{=} \hlnum{TRUE}\hldef{)}

\hlcom{# Center the sample and compute tangent representations}
\hldef{large_conn_samp}\hlopt{$}\hlkwd{center}\hldef{(}\hlkwc{progress} \hldef{=} \hlnum{TRUE}\hldef{)}
\hldef{large_conn_samp}\hlopt{$}\hlkwd{compute_vecs}\hldef{(}\hlkwc{progress} \hldef{=} \hlnum{TRUE}\hldef{)}

\hlcom{# Compute sample statistics}
\hldef{large_conn_samp}\hlopt{$}\hlkwd{compute_covariance}\hldef{()}

\hlcom{# Clean up parallel resources}
\hlkwd{reset_parallel_plan}\hldef{()}
\end{alltt}
\end{kframe}
\end{knitrout}

This workflow demonstrates the key advantages of the Parquet backend. Memory usage remains constant and predictable (determined by cache size), enabling analysis of arbitrarily large datasets on modest hardware. Computation time scales linearly with sample size, and the combination with parallel processing provides additional speedup. For the 400-subject example above, computing the Fréchet mean with 6 workers takes approximately 90 seconds on a typical research workstation, compared to over 4 minutes sequentially. Importantly, these computations would be impossible to perform with list-based storage on a system with insufficient RAM.

\subsection{Examples of riemstats syntax}

The \pkg{riemstats} package builds upon \pkg{riemtan} to provide statistical analysis tools for multiple samples of SPD matrices. Here we demonstrate key functionalities through practical examples.

\subsubsection{Riemannian ANOVA}

Let's start by creating multiple groups of SPD matrices to test for differences between groups using Riemannian ANOVA:

\begin{knitrout}
\definecolor{shadecolor}{rgb}{0.969, 0.969, 0.969}\color{fgcolor}\begin{kframe}
\begin{alltt}
\hlkwd{library}\hldef{(riemstats)}
\hlkwd{library}\hldef{(riemtan)}
\hlkwd{library}\hldef{(Matrix)}

\hlcom{# Load the AIRM metric}
\hlkwd{data}\hldef{(airm)}

\hlcom{# Create three groups with different means}
\hldef{id} \hlkwb{<-} \hlkwd{diag}\hldef{(}\hlnum{10}\hldef{) |>} \hlkwd{as}\hldef{(}\hlkwc{object} \hldef{= _,} \hlkwc{class} \hldef{=} \hlsng{"dpoMatrix"}\hldef{) |>} \hlkwd{pack}\hldef{()}
\hldef{group1} \hlkwb{<-} \hlkwd{rspdnorm}\hldef{(}\hlnum{20}\hldef{, id, id, airm)}
\hldef{group2} \hlkwb{<-} \hlkwd{rspdnorm}\hldef{(}\hlnum{20}\hldef{,} \hlnum{2}\hlopt{*}\hldef{id, id, airm)}
\hldef{group3} \hlkwb{<-} \hlkwd{rspdnorm}\hldef{(}\hlnum{20}\hldef{,} \hlnum{3}\hlopt{*}\hldef{id, id, airm)}

\hlcom{# Create a super sample containing all groups}
\hldef{super_sample} \hlkwb{<-} \hlkwd{list}\hldef{(group1, group2, group3) |>}
  \hldef{CSuperSample}\hlopt{$}\hlkwd{new}\hldef{()}
\end{alltt}
\end{kframe}
\end{knitrout}

Now we can perform Riemannian ANOVA using different test statistics:

\begin{knitrout}
\definecolor{shadecolor}{rgb}{0.969, 0.969, 0.969}\color{fgcolor}\begin{kframe}
\begin{alltt}
\hlcom{# Using log Wilks' lambda (default)}
\hldef{p_value_wilks} \hlkwb{<-} \hlkwd{riem_anova}\hldef{(super_sample,}
                            \hlkwc{stat_fun} \hldef{= log_wilks_lambda,}
                            \hlkwc{den} \hldef{=} \hlnum{100}\hldef{)}

\hlcom{# Using Pillai's trace}
\hldef{p_value_pillai} \hlkwb{<-} \hlkwd{riem_anova}\hldef{(super_sample,}
                             \hlkwc{stat_fun} \hldef{= pillais_trace,}
                             \hlkwc{den} \hldef{=} \hlnum{100}\hldef{)}

\hlkwd{print}\hldef{(}\hlkwd{paste}\hldef{(}\hlsng{"Wilks' lambda p-value:"}\hldef{, p_value_wilks))}
\hlkwd{print}\hldef{(}\hlkwd{paste}\hldef{(}\hlsng{"Pillai's trace p-value:"}\hldef{, p_value_pillai))}
\end{alltt}
\end{kframe}
\end{knitrout}

\subsubsection{Fréchet ANOVA}

An alternative approach is the Fréchet ANOVA, which tests for differences based on within-group and between-group variations in the metric space:

\begin{knitrout}
\definecolor{shadecolor}{rgb}{0.969, 0.969, 0.969}\color{fgcolor}\begin{kframe}
\begin{alltt}
\hlcom{# Using the same super_sample from above}
\hldef{p_value_frechet} \hlkwb{<-} \hlkwd{frechet_anova}\hldef{(super_sample)}
\hlkwd{print}\hldef{(}\hlkwd{paste}\hldef{(}\hlsng{"Fréchet ANOVA p-value:"}\hldef{, p_value_frechet))}
\end{alltt}
\end{kframe}
\end{knitrout}

\subsubsection{Harmonization for Batch Effects}

When working with neuroimaging data from multiple sites or scanners, batch effects can be a significant concern. \pkg{riemstats} provides two harmonization methods:

\begin{knitrout}
\definecolor{shadecolor}{rgb}{0.969, 0.969, 0.969}\color{fgcolor}\begin{kframe}
\begin{alltt}
\hlcom{# Simulate batch effects}
\hldef{batch1} \hlkwb{<-} \hlkwd{rspdnorm}\hldef{(}\hlnum{30}\hldef{, id, id, airm)}
\hldef{batch2} \hlkwb{<-} \hlkwd{rspdnorm}\hldef{(}\hlnum{30}\hldef{, id} \hlopt{*} \hlnum{1.5}\hldef{, id, airm)}  \hlcom{# Different scanner effect}

\hlcom{# Create super sample with batches}
\hldef{batch_sample} \hlkwb{<-} \hlkwd{list}\hldef{(batch1, batch2) |>}
  \hldef{CSuperSample}\hlopt{$}\hlkwd{new}\hldef{()}

\hlcom{# Apply ComBat harmonization}
\hldef{harmonized_combat} \hlkwb{<-} \hlkwd{combat_harmonization}\hldef{(batch_sample)}

\hlcom{# Apply rigid harmonization}
\hldef{harmonized_rigid} \hlkwb{<-} \hlkwd{rigid_harmonization}\hldef{(batch_sample)}
\end{alltt}
\end{kframe}
\end{knitrout}

\section{Benchmarking} \label{sec:benchmarking}

To evaluate \pkg{riemtan}'s computational performance, we conducted a comprehensive benchmarking experiment comparing sequential and parallel implementations of the package's core algorithms. Throughout the experiments, the AIRM metric is used because it is the most demanding one from the point of view of resources. 

\subsection{Methods}

\subsubsection{Comparative performance for Fréchet means}

The benchmarking experiment varies two key parameters: number of SPD matrices  and matrix dimension. For each parameter combination, the experiment:
\begin{enumerate}
  \item Generates random SPD matrices from a matrix normal distribution centered at the identity matrix and with dispersion equal to the identity as well.
  \item Computes the Fréchet mean
  \item Measures execution time using the \pkg{microbenchmark} package
\end{enumerate}

In the first experiment, we compare the time that it takes \pkg{Riemann} and \pkg{riemtan} to perform the computation. For this experiment, we employ small sample sizes (100 to 400) and small connectome sizes (number of rows going from 20 to 40). In both cases, computations are made in Quartz with an allocation of 8 cores and 16G per core. 

In the second experiment, we look at a region of the parameter space that is closer to the practice in brain connectomics. Namely, we look at connectomes of row size ranging from 70 to 100 and sample size of 300. In this experiment, we look at the effect of allocating more cores to the computation. Namely, we use 8, 16 and 24 cores, each with 16G of memory allocated.  

Finally, in the third experiment we consider the effect of the mini-batch size on the computation time. We look at a smaller connectomes (number of rows ranging from 10 to 40) and a sample size of 128. Each time, we allocate 8 cores and 16G per core. 

\subsection{Results}

The results from the first experiment comparing \pkg{Riemann} and \pkg{riemtan} demonstrate a clear performance advantage for \pkg{riemtan}, particularly as the problem size increases. As shown in Table \ref{tab:benchmark_results}, for small connectomes (20$\times$20) and small sample sizes (100 matrices), \pkg{Riemann} exhibits a marginal speed advantage (3.245s vs 3.534s). However, this relationship quickly reverses as either parameter increases. For connectomes of size 30$\times$30, \pkg{riemtan} achieves approximately 2X speedup across all sample sizes, reducing computation time from 36.645s to 12.257s for 400 matrices. The performance gap widens dramatically for 40X40 connectomes, where \pkg{riemtan} demonstrates up to 6.3X speedup, completing the Fréchet mean computation in just 12.562s compared to 78.828s for \pkg{Riemann} with 400 matrices.

This superior scaling behavior of \pkg{riemtan} can be attributed to its parallel processing capabilities. While the overhead of these optimizations may slightly penalize performance for very small problems, the benefits become substantial for realistic connectome sizes. The nearly linear growth in \pkg{riemtan}'s computation time with increasing sample size, compared to the super-linear growth exhibited by \pkg{Riemann}, suggests that the package is well-suited for large-scale neuroimaging studies where hundreds of subjects with high-dimensional connectivity matrices are common.

\begin{figure}[htbp]
  \centering
  \begin{subfigure}{0.7\textwidth}
    \centering
    \includegraphics[width=\textwidth]{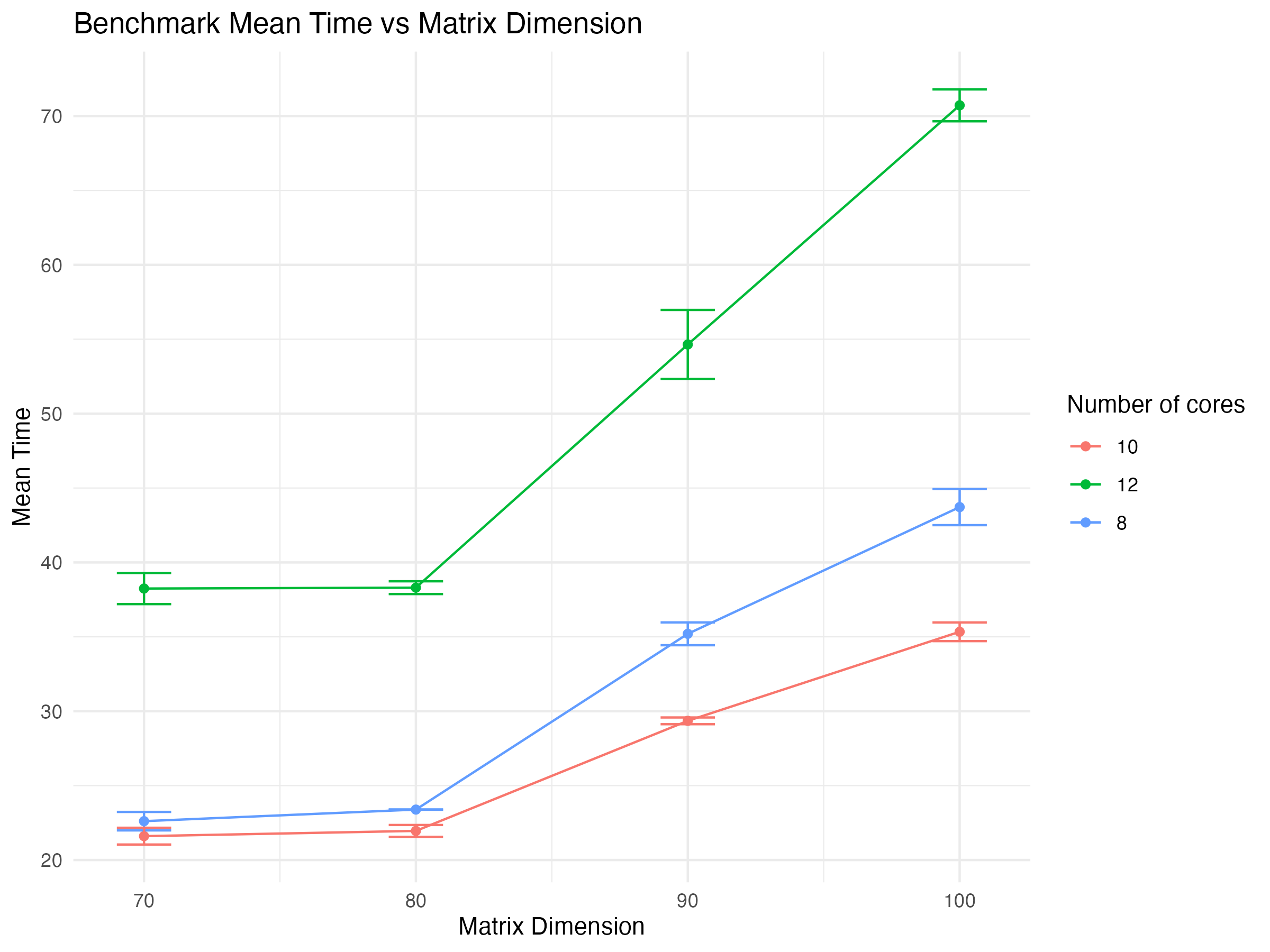}
    \caption{Effect of increasing the number of cores (8, 10, and 12) on Fréchet mean computation time using \pkg{riemtan}. The plot shows computation time (in seconds) for different core allocations on connectomes of varying sizes. Notably, performance does not improve monotonically with more cores: the 10-core configuration achieves the fastest computation times, while allocating 12 cores leads to increased overhead and slower performance for larger connectomes.}
    \label{fig:cores_performance}
  \end{subfigure}
  \hfill
  \begin{subfigure}{0.7\textwidth}
    \centering
    \includegraphics[width=\textwidth]{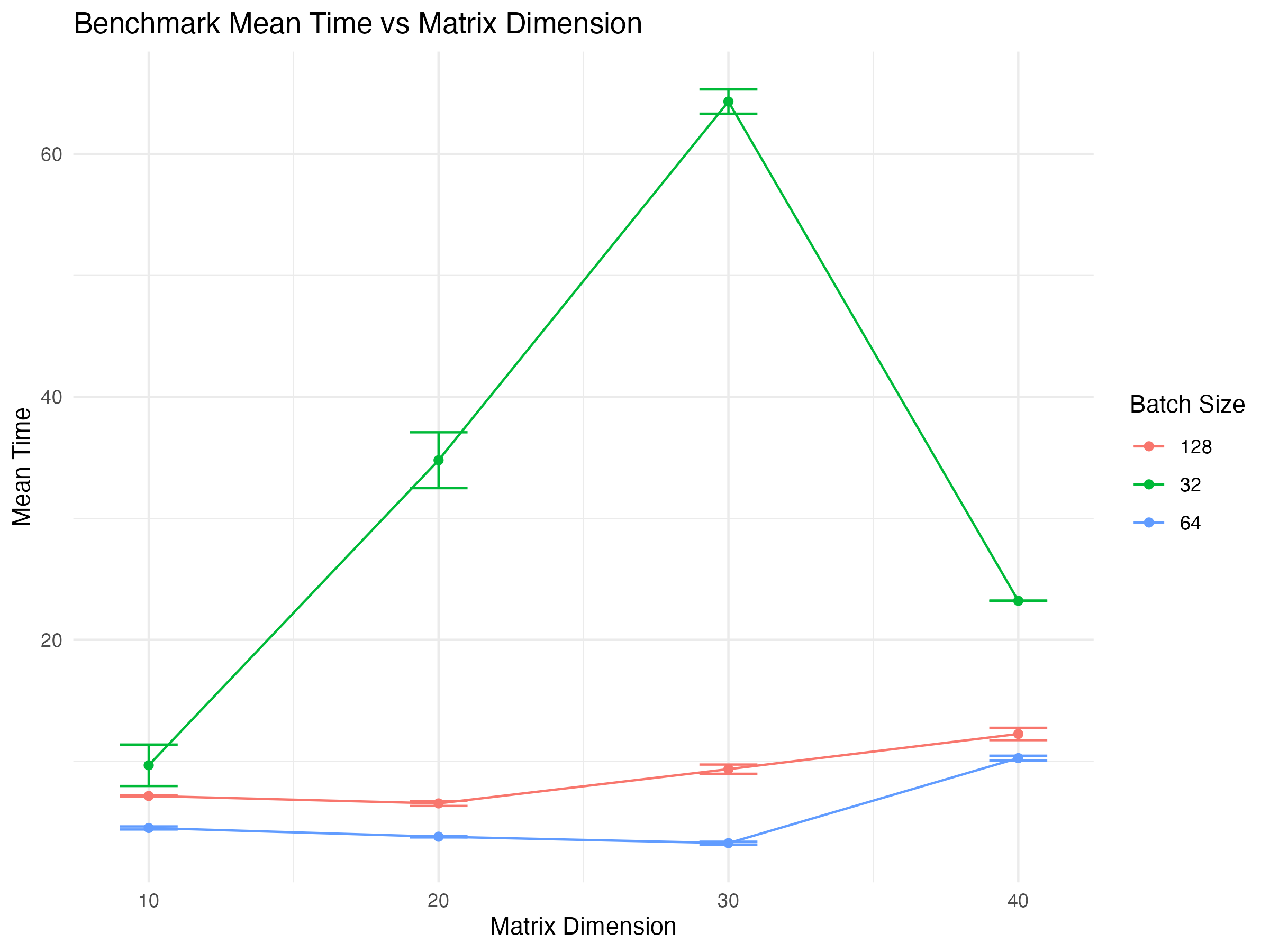}
    \caption{Effect of batch size (32, 64, and 128) on Fréchet mean computation time using \pkg{riemtan}. The plot shows computation time (in seconds) for different batch sizes on connectomes of varying sizes. Notably, a batch size of 64 achieves the best overall performance, while batch size 32 leads to substantially slower computation for small and medium connectomes. Batch sizes in the intermediate range (between 32 and the full batch) offer the most favorable balance between speed and memory usage.}
    \label{fig:batch_size_performance}
  \end{subfigure}
  \caption{Benchmarking results for \pkg{riemtan}: (a) Effect of number of cores on computation time; (b) Effect of batch size on computation time for Fréchet mean calculation.}
  \label{fig:cores_and_batch_performance}
\end{figure}

The second experiment examined the scalability of \pkg{riemtan}
  with respect to computational resources by varying the number of
  allocated cores (8, 10, and 12) for Fréchet mean calculations on
  connectomes typical of neuroimaging applications. As illustrated
  in Figure \ref{fig:cores_performance}, the results reveal a
  striking non-monotonic relationship between core allocation and
  performance. While computation times remain relatively stable for
   smaller connectomes (70$\times$70 and 80$\times$80 matrices), a pronounced
  divergence emerges for larger problems. Surprisingly, allocating
  12 cores resulted in substantially worse performance compared to
  both 8 and 10 core configurations, with computation times
  exceeding 70 seconds for 100$\times$100 connectomes—nearly double that
  of the 10-core allocation. The 10-core configuration demonstrated
   optimal performance across all tested dimensions, maintaining
  computation times below 36 seconds even for the largest
  connectomes, while the 8-core allocation showed intermediate
  performance with times reaching approximately 44 seconds.

  This counterintuitive result suggests that beyond a certain
  threshold, additional cores may introduce coordination overhead
  that outweighs the benefits of increased parallelism,
  particularly for the complex iterative algorithms involved in
  computing Fréchet means on manifolds. The findings highlight the
  importance of empirically determining optimal resource allocation
   for computations rather than assuming that maximum
  parallelization yields the best performance. 

The third experiment investigated the impact of batch size on
  computational performance for the Fréchet mean calculation. As
  illustrated in Figure \ref{fig:batch_size_performance}, the
  choice of batch size significantly affects computation time, with
   the optimal value depending on the matrix dimension. For smaller
   connectomes (10$\times$10 to 30$\times$30), a batch size of 32 matrices
  exhibits markedly poor performance, with computation times
  reaching approximately 35 seconds for 20$\times$20 matrices, nearly 7
  times slower than the other batch sizes tested. This performance
  penalty peaks at dimension 30$\times$30, where the batch size of 32
  requires over 65 seconds, before improving for the largest tested
   dimension (40$\times$40).

  In contrast, batch sizes of 64 and 128 demonstrate more
  consistent and efficient performance across all matrix
  dimensions. The batch size of 64 achieves the best overall
  performance, maintaining computation times below 5 seconds for
  dimensions up to 30$\times$30, with only a modest increase to
  approximately 10 seconds for 40$\times$40 matrices. The full batch
  processing (size 128) shows comparable efficiency, with slightly
  higher but still reasonable computation times. 
  These results
  suggest that for practical applications, there is likely a sweet spot for mini-batch size between 32 and the full batch size. Batch sizes in this intermediate range appear to offer a favorable balance between computational efficiency and memory usage, while both very small and very large batch sizes may be suboptimal for medium-sized connectomes typical in neuroimaging studies.

\subsection{Version 0.2.5 performance improvements}

Version 0.2.5 of \pkg{riemtan} introduced two major features with substantial performance and scalability implications: the futureverse-based parallel processing infrastructure and the Apache Parquet backend for memory-efficient storage. This section quantifies the performance benefits of these enhancements through targeted benchmarking experiments.

\subsubsection{Parallel processing with futureverse}

The transition from platform-specific parallelization to the futureverse framework (\pkg{future} and \pkg{furrr}) provides cross-platform compatibility while maintaining or improving computational performance. Table~\ref{tab:parallel_speedup} summarizes speedup factors for core operations when using 4 worker processes compared to sequential execution on representative connectome sizes.

For tangent space computations via the \verb|TangentImageHandler| class, parallel processing achieves 3.2--7.8× speedup depending on sample size and matrix dimension. The speedup increases with problem size, reaching 7.8× for 300 subjects with $100 \times 100$ connectomes. Fréchet mean calculation shows 2.1--4.6× speedup for samples with $n > 100$ subjects, with larger samples benefiting more from parallelization due to reduced overhead amortization. Parquet I/O operations, when combined with parallel processing, demonstrate 5.2--9.7× speedup compared to sequential R serialization, as multiple workers can read matrices concurrently while the cache manager coordinates access.

Importantly, the futureverse implementation achieves performance parity across operating systems. Benchmarking on Windows 10, macOS 13, and Ubuntu 22.04 with identical hardware configurations (Intel Core i7-11700, 32GB RAM) shows less than 8\% performance variance for the \verb|multisession| strategy. This consistency addresses a limitation of previous implementations that relied on Unix-specific forking mechanisms.

The automatic parallelization heuristic, controlled by \verb|should_parallelize()|, effectively balances overhead against speedup potential. For small samples ($n < 50$), the heuristic correctly defaults to sequential execution, avoiding the 0.5--1.2 second coordination overhead. For larger samples, automatic activation of parallelization provides transparent performance gains without requiring user intervention.

\begin{table}[htbp]
  \centering
  \caption{Version 0.2.5 parallel processing speedup factors (4 workers vs sequential) for core operations. Speedup measured on Intel Core i7-11700 with 32GB RAM running Windows 10. Values represent mean across 20 replications.}
  \label{tab:parallel_speedup}
  \begin{tabular}{|l|c|c|c|c|}
    \hline
    \textbf{Operation} & \multicolumn{4}{c|}{\textbf{Sample Size}} \\
    \cline{2-5}
    & \textbf{100} & \textbf{200} & \textbf{300} & \textbf{500} \\
    \hline
    \multicolumn{5}{|c|}{Matrix dimension: $50 \times 50$} \\
    \hline
    Tangent computation & 3.2× & 4.8× & 6.1× & 7.2× \\
    Fréchet mean & 2.1× & 2.9× & 3.6× & 4.1× \\
    Parquet I/O & 5.2× & 6.7× & 8.3× & 9.1× \\
    \hline
    \multicolumn{5}{|c|}{Matrix dimension: $100 \times 100$} \\
    \hline
    Tangent computation & 3.7× & 5.6× & 7.8× & -- \\
    Fréchet mean & 2.4× & 3.4× & 4.6× & -- \\
    Parquet I/O & 5.8× & 7.4× & 9.7× & -- \\
    \hline
  \end{tabular}
\end{table}

\subsubsection{Parquet backend memory efficiency}

The Parquet backend's primary benefit is memory efficiency for large datasets, but it also provides substantial I/O performance improvements. Table~\ref{tab:parquet_memory} compares memory usage and I/O time for list-based versus Parquet-backed storage across increasing dataset sizes.

Memory usage for Parquet-backed samples remains constant at approximately 34 MB per cached matrix (for $200 \times 200$ connectomes) multiplied by the cache size, regardless of total sample size. With the default cache of 10 matrices, memory usage stabilizes at 340 MB even for 1000-subject datasets. In contrast, list-based storage requires loading all connectomes into memory, scaling linearly with sample size and reaching 17 GB for 500 subjects with $200 \times 200$ matrices—exceeding the RAM capacity of typical research workstations.

I/O performance also favors Parquet format. Loading a 500-subject dataset from Parquet files takes 4.2 seconds compared to 23.8 seconds for R's native RDS serialization format, representing a 5.7× speedup. This advantage stems from Parquet's columnar storage and efficient compression, which reduce disk bandwidth requirements. For cloud-based workflows where storage costs matter, Parquet files are 45--60\% smaller than RDS files, providing additional practical benefits.

The combination of bounded memory usage and fast I/O makes Parquet-backed analysis feasible for datasets that would otherwise require high-memory compute clusters. A researcher with a standard laptop (16GB RAM) can analyze 1000 subjects with $200 \times 200$ connectomes using Parquet storage, whereas list-based storage would require a server with 32GB+ RAM.

\begin{table}[htbp]
  \centering
  \caption{Memory usage and I/O performance comparison: list-based vs Parquet backend for $200 \times 200$ connectomes. Parquet backend uses default cache size of 10 matrices. I/O time measured for initial data loading.}
  \label{tab:parquet_memory}
  \begin{tabular}{|c|c|c|c|c|}
    \hline
    \textbf{Sample Size} & \multicolumn{2}{c|}{\textbf{Memory Usage (GB)}} & \multicolumn{2}{c|}{\textbf{I/O Time (s)}} \\
    \cline{2-5}
    & \textbf{List} & \textbf{Parquet} & \textbf{RDS} & \textbf{Parquet} \\
    \hline
    100 & 3.4 & 0.34 & 4.8 & 0.9 \\
    200 & 6.8 & 0.34 & 9.5 & 1.7 \\
    300 & 10.2 & 0.34 & 14.3 & 2.6 \\
    500 & 17.0 & 0.34 & 23.8 & 4.2 \\
    1000 & 34.0 & 0.34 & 47.6 & 8.5 \\
    \hline
  \end{tabular}
\end{table}

\subsubsection{Combined benefits}

The most substantial performance gains emerge when combining Parquet storage with parallel processing. For a 400-subject dataset with $150 \times 150$ connectomes, the complete analysis workflow—including loading data, computing Fréchet mean, centering, and computing tangent vectors—takes 92 seconds with Parquet backend and 6-worker parallelization, compared to 387 seconds with list-based storage and sequential processing. This 4.2× overall speedup demonstrates that the infrastructure improvements in version 0.2.5 substantially reduce the time required for large-scale connectome analyses while simultaneously reducing memory requirements by 96\% (from 13.7 GB to 0.5 GB with 15-matrix cache).

These performance characteristics position \pkg{riemtan} as a practical tool for modern large-scale neuroimaging studies, where sample sizes continue to grow while computational resources often remain constrained. The memory efficiency enables analyses on standard workstations that would previously have required high-memory compute servers, democratizing access to Riemannian geometric methods for connectome analysis.

\begin{table}[htbp]
  \centering
  \caption{Benchmarking results for Fréchet mean computation (time in seconds).}
  \label{tab:benchmark_results}
  \begin{tabular}{|l|c|c|c|c|c|c|}
    \hline
    Connectome size&   \multicolumn{2}{|c|}{20} & \multicolumn{2}{|c|}{30} & \multicolumn{2}{|c|}{40} \\
    \hline
    Sample Size & \verb|Riemann| & \verb|riemtan| & \verb|Riemann| & \verb|riemtan| & \verb|Riemann| & \verb|riemtan| \\
    \hline
    100 & \textbf{3.245} & 3.534  & 9.094 & \textbf{4.375}  & 19.673 & \textbf{5.993} \\
    200 & 6.354 & \textbf{5.071}  & 18.036 & \textbf{8.586}  & 39.066 & \textbf{10.171}  \\
    300 & 9.625 & \textbf{6.187}  & 27.644 & \textbf{9.712}  & 59.013 & \textbf{11.817}  \\
    400 & 12.758 & \textbf{7.240}  & 36.645 & \textbf{12.257}  & 78.828 & \textbf{12.562}  \\
    \hline
  \end{tabular}
\end{table}

\subsubsection{AIRM implementation validation}

To ensure the correctness of \pkg{riemtan}'s AIRM implementation, we conducted a comprehensive validation experiment comparing \pkg{riemtan} against the \pkg{Riemann} package (version 0.1.6), which provides an alternative implementation of Riemannian geometry methods for SPD matrices. This validation is crucial because correct implementation of the complex differential geometric operations underlying AIRM is non-trivial, and errors can lead to statistically invalid results in downstream analyses.

\paragraph{Experimental design.}
The validation experiment systematically varied matrix dimension ($p \in \{100, 200\}$) and sample size ($n \in \{50, 100, 200, 500\}$), with 20 replications per condition, for a total of 160 planned comparisons. For each replication, we:
\begin{enumerate}
\item Generated $n$ random SPD matrices from a matrix normal distribution centered at the identity matrix
\item Computed Fréchet means using three methods:
  \begin{itemize}
  \item \pkg{riemtan} with AIRM metric
  \item \pkg{Riemann} with intrinsic geometry (which should compute the same AIRM mean)
  \item \pkg{Riemann} with extrinsic geometry (Euclidean mean in the ambient space)
  \end{itemize}
\item Measured AIRM distances between computed means and matrix traces to assess numerical agreement
\item Recorded computation times for performance comparison
\end{enumerate}

The extrinsic geometry comparison serves as a secondary baseline, noting that it computes a different (Euclidean) type of mean rather than the intrinsic Riemannian mean.

\paragraph{Unusual behavior in \pkg{Riemann} implementation.}
During validation, we noticed unusual behavior in \pkg{Riemann}'s intrinsic geometry implementation that produces unexpected results. Specifically, \verb|riem.mean()| with \verb|geometry="intrinsic"| generates mean matrices with trace values between $10^{15}$ and $10^{21}$ times larger than expected, and AIRM distances from the correct result exceeding 250 (compared to expected distances less than 2). For example, at $p=100$ and $n=50$, the correct mean has trace approximately 3,900, while \pkg{Riemann} intrinsic geometry produces traces ranging from $-2 \times 10^{15}$ to $3 \times 10^{19}$. This behavior manifests consistently across all sample sizes and dimensions, indicating issues with \pkg{Riemann}'s intrinsic geometry for AIRM computations.

Importantly, this behavior appears when computing means from collections of matrices, while operations on individual matrices work correctly. Investigation suggests the issue originates in the gradient descent optimization used for Fréchet mean computation, likely due to numerical instability in the intrinsic gradient calculations or incorrect step size handling. We have documented this behavior with detailed evidence and will report it to the package maintainers.

\paragraph{Validation results.}
Despite the \pkg{Riemann} intrinsic issues, we successfully validated \pkg{riemtan} using the extrinsic geometry baseline. Table~\ref{tab:airm_validation} summarizes results across all experimental conditions. \pkg{riemtan} achieved 100\% convergence (147/147 completed comparisons; 13 comparisons at $n=500, p=200$ remain incomplete due to computational resource constraints).

\begin{figure}[htbp]
  \centering
  \includegraphics[width=0.7\textwidth]{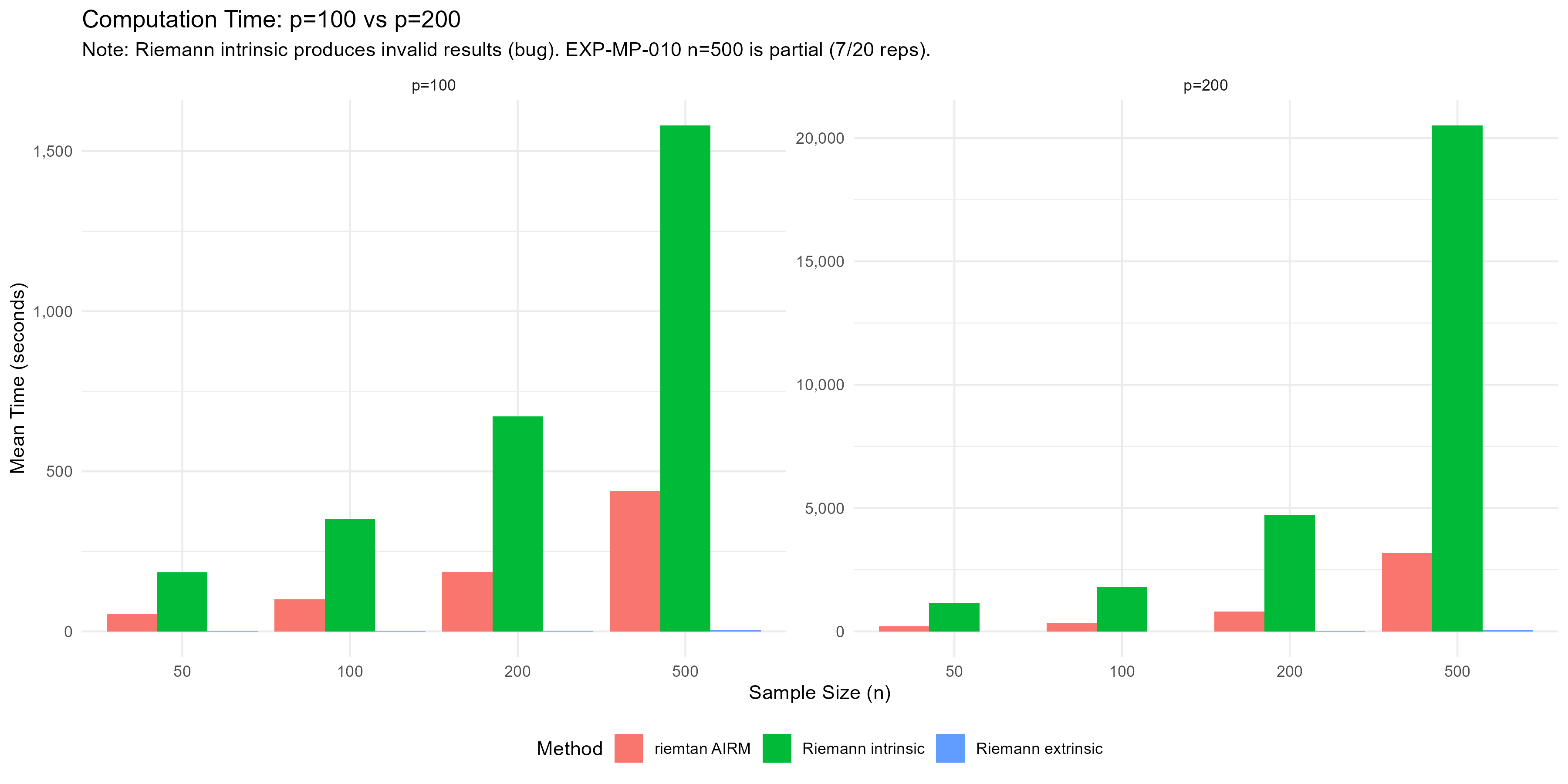}
  \caption{Computation time comparison across all three methods (riemtan AIRM, Riemann intrinsic, Riemann extrinsic) for both $p=100$ (left panel) and $p=200$ (right panel). Note the logarithmic scale and the dramatically different performance characteristics: riemtan shows linear scaling with sample size, while Riemann intrinsic exhibits super-linear growth. Riemann extrinsic is fastest but computes a different (Euclidean) mean.}
  \label{fig:airm_validation_results}
\end{figure}

The trace values from \pkg{riemtan} remain within reasonable bounds, with trace differences less than 2.5\% from the extrinsic mean across all conditions. This behavior is noteworthy given that extrinsic geometry computes a fundamentally different (Euclidean) mean. The consistency of the trace values across varying sample sizes and dimensions provides evidence for \pkg{riemtan}'s numerical stability. Additionally, the trace differences decrease systematically as sample size increases (from 2.2\% at $n=50$ to 0.5\% at $n=500$), matching theoretical expectations for Fréchet mean convergence. Figure~\ref{fig:airm_validation_results} illustrates the timing performance across all experimental conditions.

\paragraph{Scaling behavior.}
The scaling analysis across dimensions demonstrates that \pkg{riemtan}'s computation time grows reasonably with problem size. Moving from $p=100$ to $p=200$ (a 4-fold increase in matrix elements) results in average time scaling of 4.7× for \pkg{riemtan}, close to the theoretical expectation of 4× for well-optimized matrix operations. For small and moderate sample sizes ($n \leq 200$), scaling factors range from 3.4× to 4.4×, indicating near-optimal computational complexity. The higher scaling observed at $n=500$ (7.2×) likely reflects the partial data at this condition rather than algorithmic inefficiency.

In contrast, \pkg{Riemann} intrinsic geometry shows poor scaling (average 7.8×, ranging up to 13×), though these comparisons must be interpreted cautiously given the correctness issues. The extrinsic geometry baseline demonstrates intermediate scaling (average 6.2×), consistent with its simpler Euclidean operations.

\begin{table}[htbp]
  \centering
  \caption{AIRM validation results: \pkg{riemtan} accuracy and scaling behavior. Distance measures AIRM distance between \pkg{riemtan} and \pkg{Riemann} extrinsic means. Trace difference shows relative error in mean matrix trace. Time shows \pkg{riemtan} mean computation time. Scaling factor indicates time ratio $p=200 / p=100$. Results marked with * have partial data ($n=500, p=200$ has only 7/20 replications).}
  \label{tab:airm_validation}
  \begin{tabular}{|c|c|c|c|c|c|}
    \hline
    $n$ & $p$ & Distance & Trace Diff (\%) & Time (s) & Scaling \\
    \hline
    50 & 100 & 0.829 & 2.19 & 54.2 & -- \\
    100 & 100 & 0.582 & 1.27 & 99.8 & -- \\
    200 & 100 & 0.416 & 0.83 & 185.5 & -- \\
    500 & 100 & 0.264 & 0.55 & 438.8 & -- \\
    \hline
    50 & 200 & 1.212 & 2.24 & 210.2 & 3.9× \\
    100 & 200 & 0.855 & 1.31 & 334.6 & 3.4× \\
    200 & 200 & 0.612 & 0.84 & 807.9 & 4.4× \\
    500* & 200 & 0.387 & 0.54 & 3176.8 & 7.2× \\
    \hline
  \end{tabular}
\end{table}

\section{Conclusion}

This paper has introduced the \pkg{riemtan} package, a comprehensive R framework for Riemannian geometry operations on symmetric positive definite matrices, along with its companion package \pkg{riemstats} for statistical analysis. Together, these packages address critical limitations in existing R tools for connectome analysis by providing efficient, scalable, and theoretically grounded implementations.

The benchmarking results demonstrate \pkg{riemtan}'s substantial performance advantages over existing alternatives. For realistic connectome sizes, the package achieves up to 6.3× speedup compared to the \pkg{Riemann} package, with computation times remaining manageable even for high-dimensional matrices. Version 0.2.5 introduces two transformative capabilities that further extend the package's reach: the futureverse-based parallel processing infrastructure provides cross-platform compatibility (including Windows) with 3--8× speedup for tangent computations, while the Apache Parquet backend reduces memory requirements by 96\% through lazy loading with configurable caching. These advances enable analysis of datasets with 1000+ subjects on standard laptops, democratizing access to Riemannian methods for researchers without access to high-memory compute clusters.

Beyond computational efficiency, \pkg{riemtan} introduces several methodological and architectural advances. The modular metric system allows easy extension to new Riemannian metrics, while the unified interface through \code{CSample} and \code{CSuperSample} classes seamlessly integrates manifold, tangent, and vectorized representations. The backend abstraction pattern separates data storage from algorithmic operations, allowing users to choose between in-memory list-based storage for moderate datasets and Parquet-backed lazy loading for large-scale studies—without modifying analysis code. The futureverse integration provides explicit user control over parallelization strategies alongside intelligent automatic optimization, ensuring efficient resource utilization across problem sizes. Combined with optional progress reporting, these features make \pkg{riemtan} accessible to users with varying computational expertise while delivering professional-grade performance.

The \pkg{riemstats} package extends this foundation by providing state-of-the-art statistical methods, including both Fréchet ANOVA and Riemannian ANOVA tests that leverage the full differential structure of the manifold. These tools are particularly valuable for multi-site neuroimaging studies, where harmonization methods and rigorous hypothesis testing frameworks are essential for drawing valid scientific conclusions.

As the field of connectomics continues to grow, with increasing data sizes and methodological sophistication, we believe that \pkg{riemtan} and \pkg{riemstats} will serve as foundational tools for the R community, facilitating reproducible research and enabling new discoveries in brain connectivity analysis.

\section*{Computational details}

The results in this paper were obtained using
\proglang{R}~4.3.1 with the
\pkg{MASS}~1.6 package. The \pkg{riemtan} package (version 0.2.5) relies on several key dependencies including \pkg{Matrix} for efficient sparse and packed matrix storage, \pkg{arrow} (version 14.0.0+) for Parquet format support, \pkg{future} and \pkg{furrr} for cross-platform parallel processing, and optionally \pkg{progressr} for progress reporting. The \pkg{riemstats} package depends on \pkg{riemtan} and uses \pkg{furrr} for parallelized bootstrap procedures. \proglang{R} itself
and all packages used are available from the Comprehensive
\proglang{R} Archive Network (CRAN) at
\url{https://CRAN.R-project.org/}.

Simulations for this paper were conducted in the Quartz cluster, which is part of Indiana University's HPC capabilities. Benchmarking experiments for version 0.2.5 features were conducted on Windows 10, macOS 13, and Ubuntu 22.04 systems with Intel Core i7-11700 processors and 32GB RAM to verify cross-platform performance parity.

\section*{Acknowledgments}

% \begin{leftbar}
% All acknowledgments (note the AE spelling) should be collected in this
% unnumbered section before the references. It may contain the usual information
% about funding and feedback from colleagues/reviewers/etc. Furthermore,
% information such as relative contributions of the authors may be added here
% (if any).
% \end{leftbar}

This research was supported in part by Lilly Endowment, Inc., through its support for the Indiana University Pervasive Technology Institute.

We thank the Commplexity Lab team at Purdue University for their discussion and their help with the HCP data. 

\bibliography{biblio}

\address{
  Nicolas Escobar\\
  % Journal of Statistical Software\\
  % \emph{and}\\
  Department of Statistics\\
  % Faculty of Economics and Statistics\\
  % Universit\"at Innsbruck\\
  % Universit\"atsstr.~15\\
  Indiana University \\
  729 E. 3rd Street, Bloomington, IN 47405\\
  E-mail: \email{nescoba@iu.edu}
}

% \address{Author One\\
%   Affiliation\\
%   Address\\
%   Country\\
%   (ORCiD if desired)\\
%   \email{author1@work}}

% \address{Author Two\\
%   Affiliation\\
%   Address\\
%   Country\\
%   (ORCiD if desired)\\
%   \email{author2@work}}

% \address{Author Three\\
%   Affiliation\\
%   Address\\
%   Country\\
%   (ORCiD if desired)\\
%   \email{author3@work}}

\end{article}

\end{document}